\documentclass[aps,prb,showpacs,twocolumn,amsmath,amssymb]{revtex4}
\usepackage{epsfig}
\usepackage{bm}
\def\be{\begin{equation}}
\def\ee{\end{equation}}

\begin{document}
\title{Quantum Nondemolition Measurement of a Kicked Qubit}
\author{Andrew N. Jordan and  Markus B\"uttiker}
\affiliation{D\'epartement de Physique Th\'eorique, Universit\'e de Gen\`eve,
        CH-1211 Gen\`eve 4, Switzerland}
\date{June 22, 2004}
\begin{abstract}
We propose a quantum nondemolition measurement using a kicked
two-state system (qubit).  By tuning the waiting time between kicks to
be the qubit oscillation period, the kicking apparatus performs a
nondemolition measurement.  While dephasing is unavoidable, the
nondemolition measurement can (1) slow relaxation of diagonal density
matrix elements, (2) avoid detector back-action, and (3) allow for a
large signal-to-noise ratio.  Deviations from the ideal behavior are
studied by allowing for detuning of the waiting time, as well as
finite-time, noisy pulses.  The scheme is illustrated with a
double-dot qubit measured by a gate-pulsed quantum point contact.
\end{abstract} 
\maketitle

\section{Introduction} 
Motivated by general interest in mesoscopic quantum  
phenomena, as well as by specific interest in quantum computation, 
there has been intensive experimental and theoretical effort 
to study coherence properties and measurement possibilities of 
solid state qubits. 
In particular, there has been great interest in different  
quantum measurement proposals. 
This paper is concerned with a type of weak measurement, referred to as 
a Quantum Nondemolition (QND) measurement. 
QND measurements are important when one wishes to extract
information about a quantum state over a period of time without 
destroying it.  This type of measurement has been
proposed and implemented in optics, \cite{optics1,optics2}
gravity-wave antennas, \cite{gravity} and various realizations
of harmonic oscillators \cite{ho}. 
QND measurements have also been proposed as a quantum computing
measurement method for single spins. \cite{brun,spin}

A weak measurement of a quantum system occurs when 
the state of the system is not perfectly correlated  
with the detector, in contrast to the usual projective 
measurement.  Weak measurement has an advantage over projective 
measurement in that one can study the coherent oscillations  
of a single quantum state over a long time period, whereas with 
projective measurements, one must repeat the measurement many times to 
see the quantum coherent oscillations. 
However, weak measurements have the disadvantage that they usually  
act back on the system, so that one is never sure whether  
the result stems from the original quantum state alone. 
QND schemes try to overcome this limitation 
by coupling the detector to the system via an 
operator that represents an approximate constant of motion of the  
quantum system.  In this way, the detector disturbs the system in a 
minimal way. 

\begin{figure}[t,b,h] 
\label{fig1} 
\epsfxsize=2in 
\center{\epsfbox{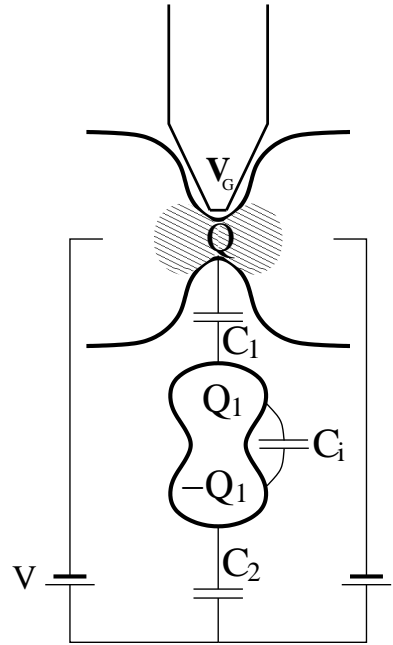}} 
\caption{The quantum nondemolition measurement performed with a
  quantum point contact.  A double-dot with charge $Q_1$ is
  capacitively coupled to a quantum point contact.  Current through
  the quantum point contact is periodically switched on and off with
  the help of a gate voltage $V_G$ with the waiting time between
  pulses given by the Rabi oscillation period of the double dot.}
\end{figure} 
Recently, Averin proposed a QND scheme for a two-state system that
involves a time dependent coupling which follows the qubit in
spin-space while weakly measuring it.\cite{averin3} The scheme was
illustrated with a Josephson-junction system as an example of a
mesoscopic qubit.  The purpose of this paper is to generalize Averin's
qubit nondemolition measurement.  The method we propose is quite
general and could be implemented in many different detector
realizations.  The basic idea is that rather than try to chase the
spin around in spin space, we let it come to us.  This is done by
strobing the qubit/detector coupling constant in time, so the detector
only has access to the qubit at approximately discrete moments.  In
this reduced problem, by fixing the waiting time between pulses to be a
multiple of the qubit oscillation time $\tau_q$, the observable we
want to measure is static in time.  Thus, by measuring over many
pulses, a large (back-action free) signal-to-noise ratio may be
obtained.  The liability of this proposal is that one cannot detect
what is happening while the coupling is turned off, so the qubit must
have a reasonably high quality factor.  However, this
scheme has several advantages over the original proposal: (1) the
coupling to the qubit needs only one qubit operator (the one we want
to measure), (2) an arbitrary qubit Hamiltonian is allowed, (3) in the
ideal case, no assumptions need to be made about the density matrix of
the qubit, and (4) in the non-ideal case, the pulses may be noisy.  These
advantages make this kicked QND measurement feasible with existing
technology.
 
Although kicking the qubit in time sounds rather violent, it is in
fact coherent because we choose the pulses to be
fast, so they may be described by a series of delta function
kicks. Dephasing is induced by the fluctuating measurement apparatus.
The inspiration for the present paper is from a venerable model of
quantum chaos, the kicked rotor.\cite{rotor1,rotor2} Here, the
situation is even simpler because the kicked quantum system has only
two states.  Nevertheless, it is the play-off between the waiting time
between kicks and the system dynamics that yields the interesting
physics.  In this way, the kicked QND measurement is similar to kicked
atoms which give rise to accelerator modes\cite{fishman1} or quantum
resonances.\cite{fishman2} However, the idea of ``stroboscopic
measurement'' is actually quite old, and the basic idea appears
already for harmonic oscillators in
Refs.~\onlinecite{strobe1,strobe2}, and has recently been
proposed for the nanomechanical resonator. \cite{resonator}

An implementation of the quantum nondemolition experiment is shown in
Fig.~1.  A double quantum-dot representing a charge qubit is
capacitively coupled to a quantum point contact (QPC).  The QPC is
opened and closed with the help of a plunger gate with voltage $V_G$.
Before considering this specific physical system, we first present
general theoretical considerations.  After the detector assumptions are
reviewed in Sec.~\ref{da}, the Hamiltonian is introduced and the ideal
quantum nondemolition measurement is described in Sec.~\ref{is}.
Deviations are then allowed in Secs.~\ref{de} and \ref{np} with detuning
between the detector pulses and the qubit oscillation time, as well as
noisy pulses of finite width.  We then describe the details of the
quantum double-dot measured by the QPC in Sec.~\ref{imp} (see Fig.~1).
Sec.~\ref{conc} contains our conclusions, and Appendix \ref{pm}
contains our perturbation method.

\section{Detector Assumptions}
\label{da} 
A realistic, effective detector has two necessary characteristics: the
measurement output should behave classically and the response time of
the detector should be much shorter than the period of the qubit
oscillations.  As mentioned in the introduction, there can be strong
detectors whose states are well correlated with the state of the
system and thus make a projective measurement of the qubit's state, or
there can be weak measurement which only extracts partial information
about the quantum state.  For the weak measurement, we assume that
linear response applies. \cite{averin1}
Quantitatively, if the bare detector variable is $f$, (which we treat
as a classical random variable), coupled to the qubit with
the operator $\sigma_z$, and the detector output is ${\cal O}$,
then it is given by a noisy, random part $q$, plus a
small component that reflects the dynamics of the
qubit,\cite{averin1,stone}
\be 
{\cal O}(t) = q(t) + \frac{\lambda}{2} \sigma_z(t), 
\label{deto} 
\ee 
where $\lambda$ is the response coefficient which indicates 
how sensitive the detector is, and $\sigma_z$ is the qubit
operator that we are trying to measure. 
With these assumptions, the temporal correlation
function of the detector's output is 
\begin{eqnarray}
C(t)&=&\langle {\cal O}(\tau+t) {\cal O}(\tau)\rangle_\tau  
=  2 \pi S_q \ \delta(t-0) +  
\frac{\lambda^2}{2}  K(t), \nonumber \\
K(t)&=&(1/2){\rm Tr}[\rho\, \sigma_z(t) \sigma_z(0)],
\label{c} 
\end{eqnarray}
where $S_q$ is the spectral weight of the noise $q$,
the density matrix of the qubit is $\rho$,
and we have made use of the fact that the detector response
time is much faster than the qubit oscillation time.
Cross terms of the form $\langle q \ \sigma_z\rangle $ 
vanish because the detector noise is 
uncorrelated with the qubit oscillations.
The spectral density of the detector is now given by 
\be 
S(\omega)= \int_{-\infty}^\infty \frac{dt}{2\pi} e^{i \omega t}\, C(t) =
S_q +\frac{\lambda^2}{2} \int_{-\infty}^\infty 
\frac{dt}{2 \pi}\ e^{i \omega t} K(t). 
\label{sw}
\ee 
This is the object that is experimentally
accessible for low frequencies.
If the detector is coupled to the qubit statically, 
over several oscillations the measurement device will measure the desired 
quantum operator ($\sigma_z$), but there will be detector back-action 
altering the qubit dynamics.  This back-action imposes a fundamental 
limit on the signal-to-noise ratio of 4.\cite{averin2} 
We turn now to the pulsed scheme to overcome this limitation. 

Instead of following the spin, we let it complete the Rabi oscillation
 and couple only after a full rotation.  The time scale of the on/off
 transition should be slow to the detector (so linear response still
 applies), but fast to the qubit.  If the effective qubit has higher
 energy states, then the time scale of the kick must also be
 sufficiently slow so as not to excite these states.  The dynamic
 coupling will result in a ladder of pulses at the measurement output
 device, typically of height $\lambda$.  The measurement of the
 quantum state of the qubit is extracted by examining how the rungs of
 the ladder are correlated with each other above the white noise
 background.

\section{The Ideal Scenario} 
\label{is}
We first consider the simplest case where the duration of the pulse
is zero.  The qubit is then coupled to the detector by periodic delta
functions in time of waiting time $\tau$, with $f$ describing the random
detector variable.  The Hamiltonian for this ideal scenario is given
by 
\be H= \frac{\epsilon}{2}\, \sigma_z +\frac{\Delta}{2}\, \sigma_x
+\sum_{n=-\infty}^\infty \delta(t-n\tau) \frac{\sigma_z f(t)}{2} +H_f,
\label{ham} 
\ee 
where $H_f$ is the detector Hamiltonian.
The evolution operator corresponding to this time-dependent Hamiltonian
may be found from
\be 
{\cal U}^t = {\rm T}\exp\left[-i \int_0^t dt' H(t')\right] ,
\label{u} 
\ee where T is the time ordering operator.  The time integration gives
the free propagator when the delta function is zero, while during the
kick, we can integrate the delta function over a small time interval
$\varepsilon$ to find the qubit propagator for the kick: 
\begin{eqnarray}
&& \lim_{\varepsilon\rightarrow 0}\ \exp\left\{-i
\int_{\tau-\varepsilon}^{\tau+\varepsilon} dt' \left [H_q+
\delta(t-\tau) \frac{\sigma_z f(t)}{2}\right] \right\} \nonumber \\
&& \qquad  = \exp[-i\sigma_z f(\tau)/2] .
\label{inth} 
\end{eqnarray}
The contribution of the free propagator is proportional to
$\varepsilon$ in this interval so it vanishes when we take
$\varepsilon\rightarrow 0$, while the kick contribution stays finite.
This allows us to factorize the propagator as 
\begin{eqnarray}
{\cal U}_i &=&{\cal U_F}\, {\cal U_K}, \quad {\cal U_F} = \exp[-i \tau
(\epsilon/2) \sigma_z -i \tau (\Delta/2) \sigma_x],\nonumber \\
{\cal U_K}&=&\exp[-i \sigma_z f(\tau_i)/2], \quad {\cal U}^t \vert
\psi\rangle=\prod_{i=1}^n {\cal U}_i\vert \psi\rangle.
\label{U}
\end{eqnarray} 
Here, $n=t/\tau$ is the discrete time index.  It is important to note that
${\cal U}_i$ changes during each time step because $f$ has its own
dynamics while the coupling is turned off that is governed by $H_f$.
So far, the waiting time interval $\tau$ may be arbitrary.  In the ideal
case, we choose $\tau$ to be $\tau_q=4
\pi/\sqrt{\epsilon^2+\Delta^2}$, the oscillation period of the qubit.
Then, the free propagator is simply the identity matrix, so the total
propagator is composed of $\cal U_K$'s only.

Now that we have the propagator, all dynamical properties of the qubit
may be investigated.  The first important property is that of the
coherence of the qubit.  Decoherence effects are caused by the
randomly changing $f$ variable.  If we begin kicking with an initially
pure state $\psi=(\alpha,\beta)$ in the $z$ basis, the density matrix
changes in time as
\begin{equation} 
\rho(n) = 
\begin{pmatrix}
|\alpha|^2   
 &&  Z \alpha  \beta^\ast \cr 
 Z^\ast \alpha^\ast \beta && |\beta|^2 \cr
\end{pmatrix},
\label{rho} 
\end{equation} 
where $Z = \prod_{i=1}^n e^{i f_i}\rightarrow \exp[i n \langle f
\rangle -n\,\langle\delta f^2\rangle/2]$ by the central limit theorem.
Here, we see explicitly that the pulsed coupling keeps the diagonal
matrix elements constant in time, but there is exponential dephasing
controlled by the fluctuations in $f$ whose dephasing rate is simply
the noise of the detector variable.  It is clear that information
about the $x$ and $y$ variables is being destroyed by the measuring
apparatus, while we have carefully preserved the $z$ information of
the state.

One simple but important quantity to calculate is the expectation 
value of $\sigma_z(n)$,
\be 
\langle \sigma_z(n) \rangle = {\rm Tr} [\rho \; 
{\cal U}^t  \sigma_z {{\cal U}^t}^\dagger].
\label{sz} 
\ee 
For the ideal case, the propagator is just a function of
$\sigma_z$, so $\sigma_z$ is a constant of motion.  Tr$[\rho
\sigma_z]=\vert\alpha\vert^2-\vert\beta\vert^2$, so the expectation
value is a measure of the population that is time independent.  We may
now calculate the time correlation function of the detector output.
The discrete version of Eq.~(\ref{c}) is 
\be 
C(n) = 2 \pi S_q \, \delta_{n,0} + 
(\lambda^2/2)\, {\rm Tr}[\rho\, \sigma_z \sigma_z(n)],
\label{k} 
\ee 
where $\sigma_z(n) = {\cal U}^{-n} \sigma_z {\cal U}^{n}$.
Again, $\sigma_z$ commutes with $\cal U$, $\sigma_z^2=1$ and Tr$\rho=1$, 
so $K(n)$ never decays with $n$, because every rung on the signal
ladder is identical.
If we calculate the discrete version of the spectral function, Eq.~(\ref{sw}),
the sum diverges at $\omega=0$.  This tells us that the accumulated
signal about $\sigma_z$ may be arbitrarily larger than the noise
of the detector, $S_q$, regardless of how small the response
coefficient $\lambda$ is. 
\section{Detuning}
\label{de}
We now generalize this ideal behavior to more realistic
situations.  In this section, we let the waiting time $\tau$ be close to
the qubit oscillation time, but not equal.  This gives us a
perturbation parameter $\gamma$, where $\tau/\tau_q = 1+\gamma$.  We
take $\gamma \ll 1$ and let $\rho=(1/2)\openone$ for calculational
simplicity.  Noting the identity 
$\exp[-i (\alpha/2) {\hat n} \cdot {\bf \sigma}] 
= {\rm} \openone \cos \alpha/2 -  
i {\hat n} \cdot {\bf \sigma} \sin \alpha/2$ 
with $\alpha = \tau \sqrt{\epsilon^2+\Delta^2}$ 
and the unit vector $\hat n$ chosen to give the free propagator 
($n_z= \tau \epsilon/\alpha$, $n_x= \tau \Delta/\alpha$),  
we choose $\alpha = 4 \pi +\gamma$ implying 
${\cal U_F}=\exp[-i (\gamma/2) {\hat n} \cdot {\bf \sigma}]$. 
This is the same as free evolution by an angle $\gamma$ of our choosing. 
Time evolution is now different from the ideal case because  
the free and kick evolution operators do not commute. 
We may investigate the time correlator of $\sigma_z$ by
expanding $\cal U$ to second order in $\gamma$ to find 
\be 
K(n) = 1 -  
\left( \frac{\Delta \tau_q \gamma}{2\pi} \right)^2 \left(\frac{n}{2}  +  
{\rm Re}\sum_{j=1}^{n-1} X_j\right) , 
\label{qbitauto} 
\ee 
where the sum describes the interaction with the detector 
and the $X_j$ are recursively defined by:  
\be
X_1 = e^{i f_1},\;
X_{j}=e^{i f_{j}} (1+X_{j-1}), \; j=2,\ldots,n-1. 
\label{X} 
\ee
The $f_i$ are chosen from the same probability distribution and are
assumed to be independent, so we can average over the $f$'s using
$\langle e^{i f} \rangle = \exp[i \langle f \rangle - S/2]$ where $S=
\langle \delta f^2 \rangle$ is the dephasing rate found from
Eq.~(\ref{rho}).  The sum in Eq.~(\ref{qbitauto}) may now be carried
out to yield 
\begin{widetext}
\be 
\langle K(n) \rangle = 1 - \left(\frac{\Delta \tau_q
\gamma}{2\pi}\right)^2 \left(\frac{n}{2} + \frac{e^{-(S/2)\,(n+1)} +
e^{-S/2} (n-1) - n \,e^{-S}}{(1-e^{-S/2})^2}\right) .
\label{qbitave} 
\ee 
\end{widetext}
With the assumption $\gamma\ll 1$, Eq.~(\ref{qbitave})
contains two (dimensionless) rates, the dephasing rate, $S$, and a much smaller
``nondemolition'' rate.  It is this nondemolition rate that controls
the decay for times longer than the initial transients, 
$\langle K(n)\rangle = 1 - {\cal G}_d\, n$,
\be
{\cal G}_d = \left(\frac{\Delta \tau_q \gamma}{2\pi}\right)^2 
\left(\frac{1}{2} + \frac{1}{e^{S/2}-1}\right).
\label{gammad}
\ee
Although this
represents only the second order contribution in $\gamma$, we can
extrapolate to longer time by exponentiating, $\langle K(n) \rangle
\approx \exp[-{\cal G}_d\, n]$.  Including the noise of the detector
$S_q$, the correlation function determines the spectral density
$S(\omega) = (1/2\pi) \sum_{n=-\infty}^{\infty} C(n) \cos \omega n$ of the
detector output.  Besides the back-action, there are always other
relaxation processes that are not well controlled.  We therefore
introduce ${\cal G}_e$ as a phenomenological environmental relaxation
rate to find 
\be S(\omega) = S_q + \frac{\lambda^2}{4\pi}\frac{
\sinh({\cal G}_e + {\cal G}_d)}{\sin^2(\omega/2) +\sinh^2({\cal G}_e/2 +
{\cal G}_d/2)} .
\label{ft} 
\ee 
The $\sin^2(\omega/2)$ reflects a fundamental periodicity in the
experiment:  the effect is still there if we wait some multiple of the
oscillation period.
If we further assume that $S \ll 1$ (the detector noise is weak), 
than we can expand the decay rate ${\cal G}_d$ and divide by $\tau_q$ to obtain
the real time decay rate,
\be
\Gamma_d = \left(\frac{\Delta \gamma}{2\pi}\right)^2 
\frac{2\tau_q}{\langle \delta f^2 \rangle}.
\label{realrate}
\ee
This special case is analogous to Averin's line-shape.\cite{averin3} 
The whole point of the scenario is that Eq.~(\ref{ft})
describes the natural line-width of the qubit with detector
back-action as a tunable parameter.

\section{Noisy Pulses}
\label{np}
In experiments, the ladder of pulses will have finite width
and their own noise associated with them. 
In this section, we take these effects into account.
To model this realistic situation, we consider the following Hamiltonian: 
\be 
H=\frac{\epsilon}{2} \sigma_z + \frac{\Delta}{2} \sigma_x 
+ \frac{\tau_q}{\tau_G} U(t)\, \sigma_z + H_U .
\label{noisyH} 
\ee 
The variable $U(t)$ generalizes the delta function coupling
$f(t)\delta(t-n\tau)$ in Sec.~\ref{is} by broadening it into a
pulse that lasts for a short time $\tau_G$ with a large prefactor
$r={\tau_q}/{\tau_G}$.  While the pulse is on, the detector variable
may be decomposed into an average height with fluctuations: $U(t) = U
+ \delta U(t)$.  Like $\langle f \rangle$, $U$ simply serves to
renormalize the $\sigma_z$ prefactor, ${\bar \epsilon} =\epsilon + r
U$, and will not enter into the final results.  The fluctuations are
treated in the white noise approximation, \be \langle \delta U(t)
\rangle =0, \qquad \langle \delta U(t) \delta U(t') \rangle = \langle
\delta U^2 \rangle \delta(t-t').
\label{noise} 
\ee 
Note that although $U(t)$ has units of energy, the convention of
Eq.~(\ref{noise}) defines $\langle \delta U^2 \rangle$ to also have
units of energy.  Now, we wish to calculate the ensemble averaged
correlator $K(t)$, Eq.~(\ref{c}), for one pulse after making the
diagonal assumption for the density matrix to simplify the analysis.
The method we use is analogous to the discrete case.  The equations of
motion are straightforward,
\begin{eqnarray}
{\dot \sigma}_x &=& - [{\bar \epsilon} + r\, \delta U(t)]\sigma_y, \; 
 {\dot \sigma}_y =  [{\bar \epsilon} + r\, \delta U(t)]\sigma_x
 -\Delta \sigma_z, \nonumber \\
{\dot \sigma}_z &=& \Delta \sigma_y .
\label{eom} 
\end{eqnarray}
The time correlator Eq.~(\ref{c}) may be calculated by formally
integrating Eqs.~(\ref{eom}) in time to a given order $n$, as shown in
the appendix.  The closed algebra of the Pauli matrices allows the
solution to be formally written in terms of $n$ time ordered
integrals.  The condition $r\gg 1$ selects a certain class of terms in
the perturbation theory that may be averaged and summed to all orders
yielding 
\be \langle K(t) \rangle = 1 - \frac{\Delta^2 t}{\langle
\delta U^2 \rangle r^2} - \frac{\Delta^2}{(\langle \delta U^2 \rangle
r^2)^2} \left( e^{-\langle \delta U^2 \rangle r^2 t}-1 \right).
\label{knoise} 
\ee The basic relaxation mechanism is the same for the detuning as it
is for the noisy pulses of finite width, so Eq.~(\ref{knoise}) bears a
close similarity to Eq.~(\ref{qbitave}).  Notice also that
$\bar{\epsilon}$ is not present in the formula (to this order in
perturbation theory).  When the pulse is turned off to let the qubit
evolve on its own, it will be the bare energy $\epsilon$ that
determines the qubit time $\tau_q$.  Thus, the average variable $U$
does not appear in the results.  There are two rates in
Eq.~(\ref{knoise}). The first, 
$\Gamma^w_{\rm D} = r^2\, \langle \delta U^2 \rangle$, is an
additional contribution to the fast dephasing rate. The second is
\be
\Gamma_w =\frac{1}{r^2}\,
\frac{\Delta^2}{\langle \delta U^2 \rangle},
\label{rates} 
\ee   
which clearly corresponds to the slow nondemolition rate of the noisy pulses. 
The parameter $r^{-1}=\tau_G/\tau_q$ plays the role of 
the detuning parameter $\gamma$ in the previous section.

We now wish to combine all the phenomena discussed.  When both the
detuning $\gamma$ and the width of the pulses $r^{-1}$ are small, a double
nondemolition measurement is taking place.  In this case, we can write
(in the vicinity of $\omega=0$) an approximation to the spectral noise
of the detector as a Lorentzian,
\begin{eqnarray} 
S(\omega) &=& S_q + \frac{\lambda^2}{\pi} 
\frac{\Gamma}{\omega^2+\Gamma^2},  \quad \Gamma= \Gamma_e
+\Gamma_{d} +\Gamma_{w}, \nonumber \\ 
\Gamma_d &=& \left(\frac{\gamma}{2 \pi}\right)^2 \frac{2 \Delta^2 \tau_q}
{\langle\delta f^2\rangle},  \quad
\Gamma_w = \left(\frac{1}{r}\right)^2 
\frac{\Delta^2}{\langle \delta U^2 \rangle }.
\label{boths}
\end{eqnarray} 
Here, the total broadening width of the spectral noise, $\Gamma$, has been
decomposed into contributions from the environment, the detuning of
the pulses from resonance, and the finite width of the pulses.
 
\section{Implementation with the Quantum Point Contact/Double Dot} 
\label{imp}
To see how this nondemolition measurement will work in practice, we
illustrate it by using a double quantum dot (DD) as the qubit and the
current through the Quantum Point Contact (QPC) as the measurement
apparatus.\cite{stone,averin2,gurvitz,QPC2,QPC3,pilgram,zeno,clerk}
Additionally, a gate is placed above the QPC so that one may bias the
barrier of the electrons in the QPC (Fig.~1).  In this fashion,
we can pulse the QPC so as to let current flow through during a short
time when the gate voltage is raised above the barrier of the QPC.
This pulsing is similar to the rectangular voltage pulses applied in
the experiment of Ref.~\onlinecite{DDexp} and theoretically described
in Ref.~\onlinecite{pulsed}.  The magnitude of this current will be
sensitive to whether the electron is on the upper or lower part of the
DD.  At the detector output, there will be a current pulse train that
may be analyzed statistically.  The results of the nondemolition
measurement will be seen in the spectral density of the current shot
noise, Eq.~(\ref{boths}).

The charge response time\cite{tcmb} $\tau_{RC}$ of a QPC is very fast.
Here we assume that even the single particle dynamics of the QPC
governed by $\tau_{\rm QPC} =\hbar/(e V)$ (where $V$ is the applied
bias) is fast compared to the pulse duration $\tau_G$ which is in turn
fast compared to the Rabi oscillation period of the qubit (we now
restore the $\hbar$ for clarity).  Therefore, a good separation of
time scales may be reached for the pulse time $\tau_G$, 
\be \tau_{\rm
QPC} \ll \tau_{G} \ll \tau_q.
\label{sts}
\ee For solid state qubits, there is another constraint that true
spins do not have: more states at higher energy.  To avoid exciting the
qubit into these states, a separate constraint is that $\tau_G >
\hbar/\Delta E$, where $\Delta E$ is the energy difference to the next
excited state.  In this limit, we can use the physical setup described
by Pilgram and one of the authors.\cite{pilgram} The interaction
Hamiltonian that couples the DD qubit to the detector is $H_c = {\hat
Q}_1 {\hat Q}/C_i$ where ${\hat Q}$ is the charge in the vicinity of
the QPC, playing the role of $f$ and ${\hat Q}_1$ is charge on the DD.
The current operator to be measured takes the form of a bare current
${\hat I}_{b}$ playing the role of $q$, plus a self-consistent
screening current proportional to the the potential ${\hat U}$ of the QPC,
\be {\hat I} = {\hat I}_{b} - \chi {\hat U}, \qquad 
\chi = \frac{e^2}{2 \pi \hbar} \sum_n \frac{d T_n}{d E}\, e \vert V \vert.
\label{iop} 
\ee 
The (non-equilibrium) linear-response function $\chi$ is defined in
terms of the (large) bias $V$ across the QPC, and the transmission
probabilities of channel $n$, $T_n$.  To express this in the form of
Eq.~(\ref{deto}), ${\hat U}$ must now be written in terms of the
charge ${\hat Q}_1$ on the DD and the capacitances defined in Fig.~1,
${\hat U} = [C_p/(C_i e^2 D)] {\hat Q}_1$, where $D$ is the density of
states at the Fermi energy in the scattering region, $C_p^{-1} =
C_1^{-1}+C_2^{-1}+D^{-1}$ is the polarization capacitance, and ${\hat
Q}_1 = e\, {\hat \sigma}_z$.  The deviation of the average current is
given by 
\be \langle \Delta I(t) \rangle = \frac{\lambda}{2} \langle
\sigma_z(t) \rangle,
\label{avecur} 
\ee 
while the shot noise spectrum is given by 
\be 
S_{II}(\omega) = S_{I_b I_b} + \frac{\lambda^2}{2}  {\tilde
  K}(\omega), 
\label{S} 
\ee 
where $ {\tilde K}(\omega)$ is the Fourier transform of $K(t)$.
Thus, the response coefficient $\lambda$ is simply the
difference in current output depending on whether the electron
is in the upper or lower dot.
In terms of the system parameters, it is given by 
\be
\lambda = \frac{e C_p}{\pi \hbar C_i D}\, 
\sum_n\frac{dT_n}{dE}\,e\vert V\vert.
\label{lam}
\ee 
Physically, $\lambda$ is proportional to the difference in
transmission caused by the location of the DD electron, $\delta T =
(dT/dE)\, \delta E$, where $\delta E$ is the inverse density of states
$D^{-1}$ and $C_p/C_i$ is the (small) coupling constant.

By doing an experiment where one varies the time between pulses and 
plots either the width of the line-shape of the spectral density
or the signal-to-noise ratio as a function of the waiting time, 
the point of minimum width (or maximum signal-to-noise)
finds both the oscillation time $\tau_q$ 
of the qubit as well as the system's natural (unmodified) relaxation
rate. This is because of the fact that the width is simply a sum of  
the measurement back-action and uncontrolled environmental relaxation.

\section{Conclusions}
\label{conc} 
We have proposed a novel quantum nondemolition measurement on
two-state systems with a kicking measurement device.  By tuning the
waiting time between kicks to be the oscillation period of the qubit
dynamics, the weak measurement process does not act back on the
desired operator being measured, though it does destroy the
complimentary information in the conjugate variables.  
Deviations from this ideal scenario are taken into account 
by allowing for a detuning of the
waiting time as well as for noisy pulses of finite width.  These effects
may be described by an approximate Lorentzian line-shape, whose
width is the sum of three terms: the uncontrolled environmental
relaxation rate, a QND detuning contribution, and a QND noisy
pulse-width contribution.  This novel QND proposal was illustrated
with the quantum double-dot, measured by a gate-pulsed quantum point
contact. The results of the nondemolition experiment may be obtained
from the shot noise.

\section{Acknowledgments}

We thank D.~V. Averin and S. Pilgram for helpful discussions of their
work.  This work was supported by the Swiss National Science
Foundation.

\appendix

\section{Perturbation Method}
\label{pm}
We now show how to obtain Eq.~(\ref{knoise}) from
Eq.~(\ref{noise}) and Eq.~(\ref{eom}), considering the fact that the
duration of the pulse is much shorter than one coherent oscillation.
We use time-dependent perturbation theory by writing nested time
ordered integrals to a given order, and then take averages over all
the $\delta U$'s.  The primary object of interest is the correlator
$K(t) = (1/2) {\rm Tr}[\sigma_z(t) \sigma_z(0)]$.  At the initial
time, ${\rm Tr}[\sigma_x(0) \sigma_z(0)] ={\rm Tr}[\sigma_y(0)
\sigma_z(0)]=0$, so only the terms that involve $\sigma_z(0)$ will
contribute to the correlator $K(t)$.  We thus generate integral
expressions for $\sigma_z(t)$ that end in $\sigma_z(0)$.  As a simple
example, the correlator to second order in time is found from
\begin{widetext}
\be
\sigma_z(t) = \int_0^t dt' \Delta \sigma_y(t')+ \sigma_z(0)
= \int_0^t dt' \Delta \left\{ \int_0^{t'} dt'' \left[\left({\bar
\epsilon} + r \delta U(t'')\right) \sigma_x(t'')- \Delta \sigma_z(t'')\right]+
\sigma_y(0) \right\} + \sigma_z(0).
\label{2nd}
\ee
To compute $K(t)$, we
set all the time arguments of the $\sigma$'s to 0, and take the trace
with $\sigma_z(0)$.
Thus $K(t) = 1 - \Delta ^2 \int_0^t dt'\int_0^{t'} dt''
=  1 - \Delta ^2 t^2/2$.
To go to higher orders, the trick is to eliminate $\sigma_x$:
\be
\sigma_y(t)=  \sigma_y(0)+ \int_0^t dt' \left\{ - \Delta \sigma_z(t')
-\int_0^{t'} dt'' \left[ \left({\bar \epsilon} + r \delta U(t')\right)
({\bar \epsilon} + r \delta U(t''))+\Delta^2 \right] \sigma_y(t'')\right\}.
\label{trick}
\ee
\end{widetext}
The order of the expansion is given by the number of different time
integrals.
In general, there are many different terms to compute at each order.
However, because the time $t$ is much shorter than the
oscillation time, the dominate contribution is the one given 
by the maximal number of $r$'s, or $\delta U$'s.
At order $2n$, such terms are of the form $\Delta^2 r^{2n-2}$.
To average over the fluctuating potentials, we use Wick's theorem
to average the $2n-2$ $\delta U$'s in terms of $m=n-1$ 
pairwise contractions.
We then compute average time ordered integrals of the form 
\be
\int_{\{t's\}} \langle \delta U(t_1) \delta U(t_2)\ldots \delta
U(t_{2m}) \rangle .
\label{Us}
\ee
In general, contractions of this form have every possible 
time-ordering of two $\delta U$'s.  However, the contraction 
$\langle\delta U(t_i)\delta U(t_j)\rangle$
comes with a delta function enforcing $t_i=t_j$.
If time-ordered integrals occur where $t_i$ and $t_j$
are not in adjacent time intervals, then this term is zero under 
the integration.  The only term left will be the one where
all time arguments are time ordered.
Therefore, the integral Eq.~(\ref{Us}) is given by
$\langle\delta U^2\rangle^m t^{m+2}/(m+2)!$.
We may now sum these dominate contributions at every order
in time to obtain,
\begin{eqnarray}
&&K(t)=1-\frac{\Delta^2 t^2}{2!}+\sum_{n=2}^{\infty}
\frac{(-1)^n}{(n+1)!}\left[\langle\delta U\rangle^2 
r^2 \right]^{n-1}t^{n+1}, \nonumber \\
&&=1-\frac{\Delta^2 t}{\langle \delta U^2 \rangle r^2} -  
\frac{\Delta^2}{(\langle \delta U^2 \rangle r^2)^2}
 \left( e^{-\langle\delta U^2\rangle r^2 t}-1\right), 
\label{result}
\end{eqnarray}
which is the same as Eq.~(\ref{knoise}).

\end{document}